\documentclass[fleqn,10pt]{wlscirep}
\usepackage[utf8]{inputenc}
\usepackage[T1]{fontenc}
\usepackage{lineno}
\usepackage[binary-units]{siunitx}
\usepackage{multirow}
\usepackage{graphicx}
\usepackage{caption}
\usepackage{subcaption}
\usepackage{xspace}
\usepackage{todonotes}

\linenumbers

\title{Hierarchical storage management in user space for neuroimaging applications}

\author{Val\'erie Hayot-Sasson}
\author{Tristan Glatard}
\affil{Department of Computer Science and Software Engineering, Concordia University, Montreal, Canada}

\begin{abstract}

	Neuroimaging open-data initiatives have led to increased availability of large
  scientific datasets. While these datasets are shifting the processing
  bottleneck from compute-intensive to data-intensive, current standardized
  analysis tools have yet to adopt strategies that mitigate the costs
  associated with large data transfers. A major challenge in adapting neuroimaging
  applications for data-intensive processing is that they must be entirely
  rewritten. To facilitate data management for standardized neuroimaging tools, we
  developed Sea, a library that intercepts and redirects application read and
  write calls to minimize data transfer time. In this paper, we investigate the
  performance of Sea on three preprocessing pipelines implemented using standard toolboxes (FSL,
  SPM and AFNI), using three neuroimaging datasets of different sizes
  (OpenNeuro's ds001545, PREVENT-AD and the HCP dataset) on two high-performance
  computing clusters. Our results demonstrate that Sea provides large speedups (up to 32$\times$) when
  the shared file system's (e.g. Lustre) performance is deteriorated. When the
  shared file system is not overburdened by other users, performance is
  unaffected by Sea, suggesting that Sea's overhead is minimal even in cases
  where its benefits are limited. Overall, Sea is beneficial, even when performance gain
  is minimal, as it can be used to limit the number of files created on parallel file systems.
\end{abstract}
\begin{document}

\nolinenumbers

\flushbottom
\maketitle

\thispagestyle{empty}


  \section{Introduction}\label{sec:sea_neuro:introduction}
    
    The ever-growing surge in publicly available neuroimaging data has lead to new data
    management challenges, from storage infrastructure to application processing times.
    Many tools have been developed by the community to address various needs in 
    neuroimaging data management.
    For instance, standardized metadata formats, such as the Brain Imaging Data Standards (BIDS)\cite{gorgolewski2016brain},
    have been implemented to facilitate the sharing
    of the datasets, and data management tools such as DataLad~\cite{halchenko2021datalad} have been developed to provide 
    versioning and provenance capture of data.
    Furthermore, recent developments in neuroimaging pipelines have addressed computation time 
    limitations by adopting machine-learning approaches that are an order of magnitude faster than traditional image processing solutions~\cite{henschel2020fastsurfer,hoffmann2021synthmorph}.
    While all these solutions to Big Data-related data management exist, certain aspects, 
    such as processing-related data-transfer overheads
    have received limited attention.

    The current largest neuroimaging datasets,
    the Human Connectome Project (HCP)~\cite{HCP} and the UK
    Biobank~\cite{ukbiobank}, reach up to Petabytes of data. For datasets of this
    volume, parallel high-capacity remote storage solutions are commonly relied
    upon as the persistent storage layer during processing. 
    Despite their respective high-scalability, these remote file systems cannot typically
    compete with the performance of compute-local storage such as in-memory file systems (tmpfs) or solid-state drives (SSDs). Should the pipelines 
    need to transfer large amounts of data to these remote storage locations, significant data
    processing overheads may occur.

    Big Data in neuroimaging may exist in two formats: 1) very large files
    (e.g., BigBrain~\cite{amunts2013bigbrain}) and 2) small files pertaining to a
    population of individuals, such as those found in MRI datasets, that
    collectively amount to very large datasets. Differences in the number and
    size of files result in different data transfer overheads. Whereas large
    file I/O is predominantly impacted by the underlying storage bandwidth,
    small file I/O results in a non-negligible latency overhead. 

    Big Data frameworks have addressed the performance limitations arising from large data transfers via the
    implementation of two strategies: data locality and in-memory computing~\cite{zaharia2016apache, rocklin2015dask}.
    The collective aim of these strategies is to reduce overheads related to large data
    transfers by leveraging compute-local storage as cache.
    While these frameworks have been used
    to process neuroimaging data~\cite{rokem2021pan,thunder,boubela2016big},
    their popularity within the domain remains low due to efforts required to
    rewrite standardized neuroimaging toolboxes to utilize the frameworks.
    
    Newer neuroimaging applications leveraging popular neuroimaging pipeline engines
    also do not benefit from processing-related data management.
    Although engines such as Nipype~\cite{nipype}
    do not prohibit the use of data-management
    strategies, they do not facilitate the integration of these strategies into
    their resulting workflow. To give
    neuroimaging applications data management capabilities, the applications
    must interact with a file system or library that enable the strategies.


    
    
    Many researchers rely on one of two systems to meet their storage and
    computing needs: 1) High Performance Computing (HPC) clusters and 2) the
    cloud. Whereas the cloud simplifies data sharing and gives researchers
    access to a wide variety of infrastructures, HPC clusters are a
    cost-effective solution to accessing a wide array of resources for
    researchers. 
    
    HPC systems rely on scalable network-based parallel file systems (e.g.,
    Lustre) for storage. While such file systems offer excellent performance,
    they are shared between all the users on the cluster, meaning that users
    with data-intensive applications can effectively deteriorate the performance
    of the shared file system for all users on the cluster. Solutions for
    improving shared file system performance include
    throttling~\cite{huang2020ooops} the data-intensive workloads or
    recommending the use of Burst Buffers~\cite{bb} (e.g., reserving a compute
    node for storage or leveraging local compute storage during processing). The
    latter, however, requires that the user manages their data to and from the
    Burst Buffer, unless a Burst Buffer file system, such as
    BurstFS~\cite{burstfs} or GekkoFS~\cite{gekkofs} is installed.
    
    In order for a file system to be mounted by an HPC cluster user, it must be
    loadable without administrative privileges. As the applications are
    typically made to interact with POSIX-based file systems, the new file
    system must also be compliant to the format. One method to ensure that these
    conditions are met is through the \texttt{LD\_PRELOAD} trick. This trick is
    used to intercept select library calls and redefine their behaviour. It has
    been used in many projects and to create lightweight userspace versions of
    file systems~\cite{xtreemfs,burstfs,gekkofs}. 
    
    In this paper, we present Sea~\cite{hayot2022sea}, a data-management library designed to reduce
    transfer-related overheads in user space. Sea leverages the \texttt{LD\_PRELOAD} trick to redirect applications read and write calls to 
    local or remote storage transparently. We view Sea as a complement
    to the existing stack of neuroimaging tools and standards meant to facilitate 
    the processing of neuroimaging Big Data.   

    

    \section{Results}

    \subsection{Sea}
    
\begin{figure*}

    \centering
    \includegraphics[width=\columnwidth]{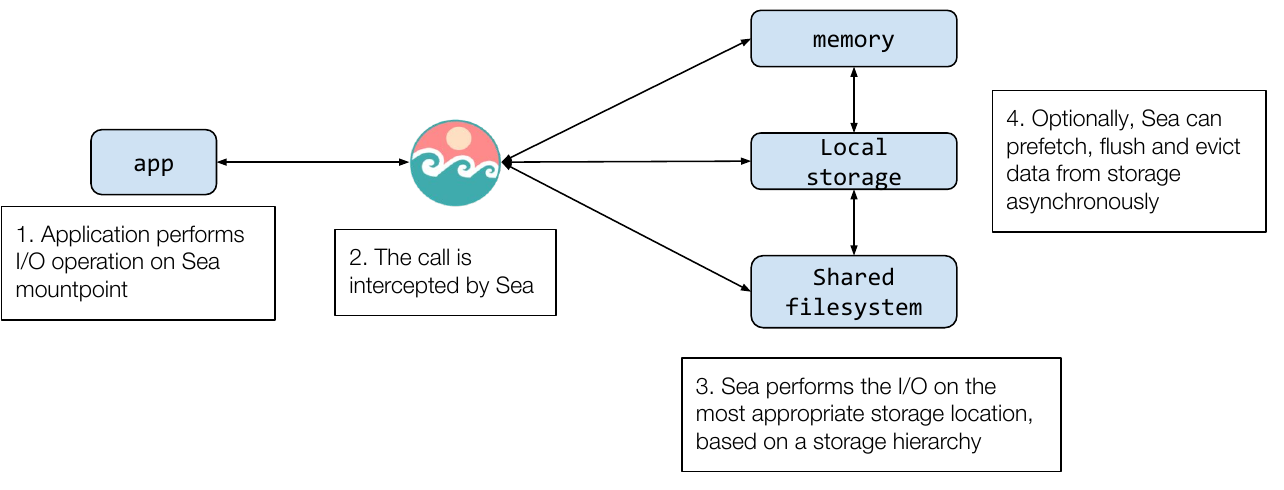}%
\caption{The Sea data-management library overview}
\label{fig:seaneuro:diagram}
\end{figure*}
    Sea (Figure~\ref{fig:seaneuro:diagram}) is a data-management library that leverages
    the \texttt{LD\_PRELOAD}
    trick to intercept POSIX file system calls, or more specifically, file access calls to the GNU C library,
    glibc, on Linux systems. This enables Sea to redirect write calls
    aimed at slower storage devices to a faster device whenever possible.
    Similarly, when intercepting read calls, Sea can choose to read from a
    faster device if a copy is available on that device. Sea decides which
    storage location it can write to based on the details provided in an
    initialization file called \texttt{sea.ini}. This file informs Sea of which
    locations it can use to read and write to, and well as their order of
    priority. Importantly, Sea is not a filesystem, but rather a
    lightweight tool that redirects files to more optimal storage devices.

    We define all locations in which Sea can redirect writes to as a cache. Such
    caches include any available storage located on the compute nodes, such as
    tmpfs (i.e. a temporary file system in virtual memory) or solid state drives (SSDs). 

    As neuroimaging pipeline results typically require post-processing and
    HPC compute-local resources are only accessible during the reserved
    duration, Sea provides functionality to flush and evict data to persistent
    shared storage. To avoid interrupting ongoing
    processing with data management operations, this is accomplished via a separate thread (known as the
    ``flusher'') that moves data from the caches to long-term storage. Users must inform Sea of files
    that need to be persisted to storage within a file called
    \texttt{.sea\_flushlist}, and temporary files which can be removed from
    cache within a file called the \texttt{.sea\_evictlist}. Both these files
    can be populated using regular expressions to denote the paths to flush and
    evict. If a file occurs in both the \texttt{.sea\_flushlist} and
    \texttt{.sea\_evictlist}, Sea will interpret this as a \texttt{move}
    operation and simply move the file from local to persistent storage rather
    than copying it. Files that will be reused by the application should only be
    flushed rather than flushed and evicted, as files can benefit from speedups
    related to reading from fast rather than slow storage. Sea currently also
    provides a rudimentary prefetch thread that can move files located within
    Sea to the fastest available cache. To use Sea's prefetch
    capabilities, a file called \texttt{.sea\_prefetchlist} needs to be
    populated using regular expressions like the flushing and eviction files.
    
    To interact with Sea, a mountpoint is used. The mountpoint
    is an empty directory that behaves as a view to all the files and
    directories stored within Sea. In order to keep track of the locations of
    files within the mountpoint, Sea mirrors the directory structure of each
    storage location across all caches. It is generally advisable to provide
    empty storage locations for Sea to use as the mirroring of large directories
    can take some time.
    
    Sea can easily be executed directly using the available containers on the
    GitHub registry, or can be compiled via source using gcc-C++ and Make. Once
    compiled, Sea can be executed using the \texttt{sea\_launch.sh} bash script.

\begin{figure*}

\begin{subfigure}{0.5\textwidth}
    \centering
    \captionsetup{width=.85\linewidth}
    \includegraphics[width=\columnwidth]{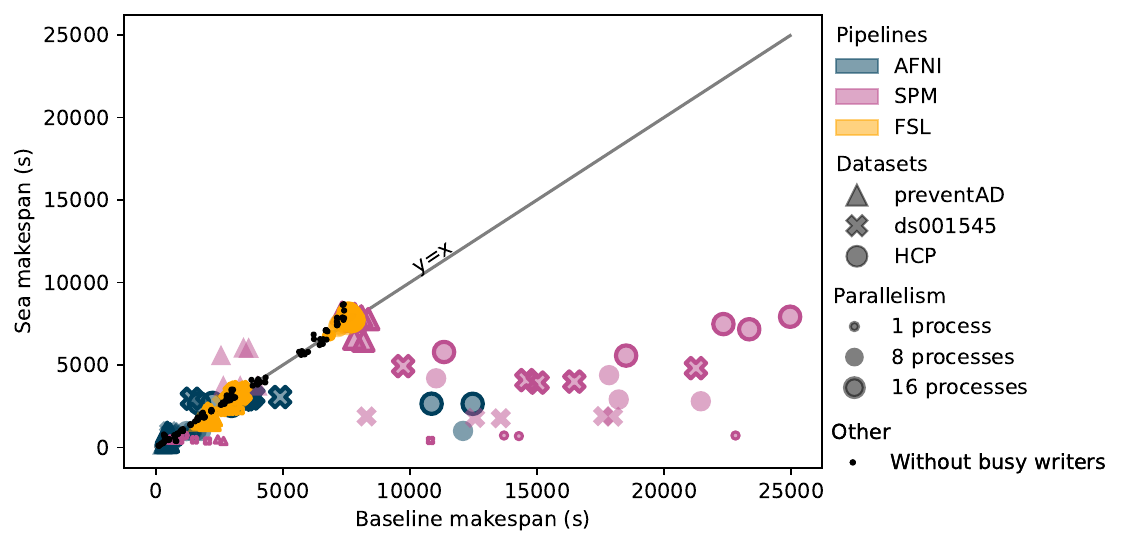}%
    \caption{Complete results}\label{fig:seaneuro:slashbinfull}
\end{subfigure}
\begin{subfigure}{0.5\textwidth}
    \centering
    \captionsetup{width=.85\linewidth}
    \includegraphics[width=\linewidth]{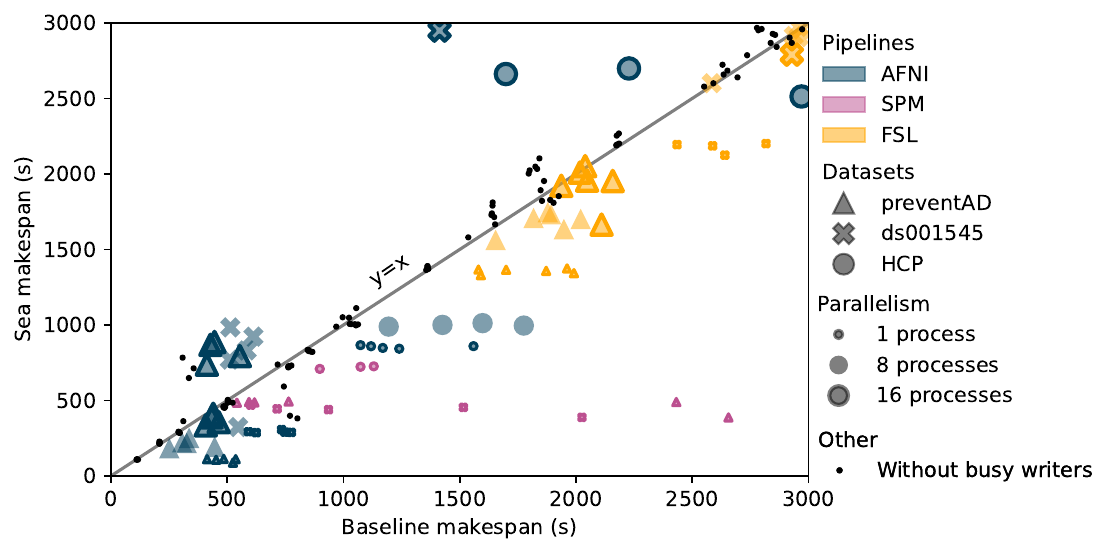}
    \caption{Zoomed between 0 and 5000 seconds}\label{fig:seaneuro:slashbinzoom}
\end{subfigure}
\caption{Makespan comparison between Sea and Baseline on controlled dedicated cluster. Makespan
denotes the total time between execution launch and completion of the last computing task.}
\label{fig:seaneuro:slashbin}
\end{figure*}

\subsection{Speedups observed in controlled environment}

We benchmarked Sea by processing 3 functional MRI datasets of increasing sizes
(ds001545: 27~GB, PREVENT-AD: 283~GB, Human Connectome Project: 79~TB, see
Table~\ref{table:sea-neuro:data}) with 3 pipelines 
from the most commonly-used neuroimaging toolboxes (AFNI, SPM, FSL), using a controlled HPC cluster 
dedicated to our experiments. We deployed ``busy writers'' in the cluster
that simulated the load from other user applications by writing to the shared file system at a controlled rate.

Speedups were obtained by using Sea when Lustre
performance had been degraded by busy writers (Figure~\ref{fig:seaneuro:slashbin}),
although performance was variable. The largest speedup across all conditions
was found to be 32$\times$ and occurred when SPM was processing a single
HCP image with 6 active busy writers.
On average, however, a single SPM pipeline
process preprocessing a single HCP image produced the greatest speedup
(12.6$\times$). The AFNI pipeline preprocessing a single fMRI image of the PREVENT-AD dataset
using a single process was the next fastest pipeline, with an average speedup of
(4.3$\times$). The FSL Feat pipeline had speedups as well, with a maximum average speedup of
1.3$\times$ when preprocessing a single PREVENT-AD image. The SPM pipeline consistently had
excellent speedups, which is likely due to a mix of prefetching the initial
input files and the I/O patterns of the application. While it was expected that the
AFNI pipeline would benefit the most from Sea due to its
minimal compute time and the amount of data it generates, it did not benefit
from Sea to the extent that the SPM pipeline did. As per Table~\ref{table:seaneuro-pipelines}, the
AFNI pipeline performs a very high number of glibc calls. While the overhead of an
individual call is likely very minimal, it is possible that the culmination of
all these calls results in a large enough bottleneck that affected both Sea and
Baseline executions to an extensive amount. Moreover, the number of glibc calls
aimed at interacting with data on the Lustre storage are very few. The AFNI pipeline still
obtained an average of 2$\times$ speedup across all experimental conditions.

The FSL Feat pipeline, in contrast, appeared to be the most compute-bound of the three
applications. Not only did it spend an extensive amount of time computing, the
amount of output data generated was least of the three. Due to the
compute-intensive nature of the FSL Feat pipeline, it is expected to have resulted in the
least significant speedups.

The HCP dataset obtained the greatest speedups with
Sea, with $2.3\times$ on average vs. $1.9\times$ for ds001545  and $1.3\times$ for PREVENT-AD.
Out of the three datasets, HCP has the largest images (see
Table~\ref{table:sea-neuro:data}). Larger individual images mean that they
occupy more page cache space and take longer to flush to Lustre. Unsurprisingly,
the next dataset with the largest images (\SI{282}{\mebi\byte} compressed for a
single image) results in the next largest speedups, followed by PREVENT-AD, the
dataset with the smallest functional images.

For both the pipelines and the datasets, we noticed that obtained speedups
decreased with increased parallelism. We did not modify the default parallelism
level of the applications in our experiments. As a result, each application
process was likely attempting to utilize all cores, resulting in longer wait times
with the increase in parallelism. With the added compute contention overhead, the benefits of using
Sea becomes more limited.

\subsection{Speedup correlated with Lustre degradation}

Baseline performance without busy writers was comparable to that of Sea's (p=0.7, two-sample unpaired t-test) as can be seen in Figure~\ref{fig:seaneuro:slashbin}. However, with
busy writers, the makespan obtained with Sea was smaller than baseline (p<1e-04, two-sample t-test). Occassional slowdowns did occur with the use of Sea. These slowdowns
may arise from the initial read of the data, as they appeared to occur less
frequently with the SPM pipeline, or due to increased CPU contention caused by Sea's rapid I/O.

\subsection{Overhead of Sea was minimal}

\begin{figure*}
\centering
\includegraphics[width=0.7\columnwidth]{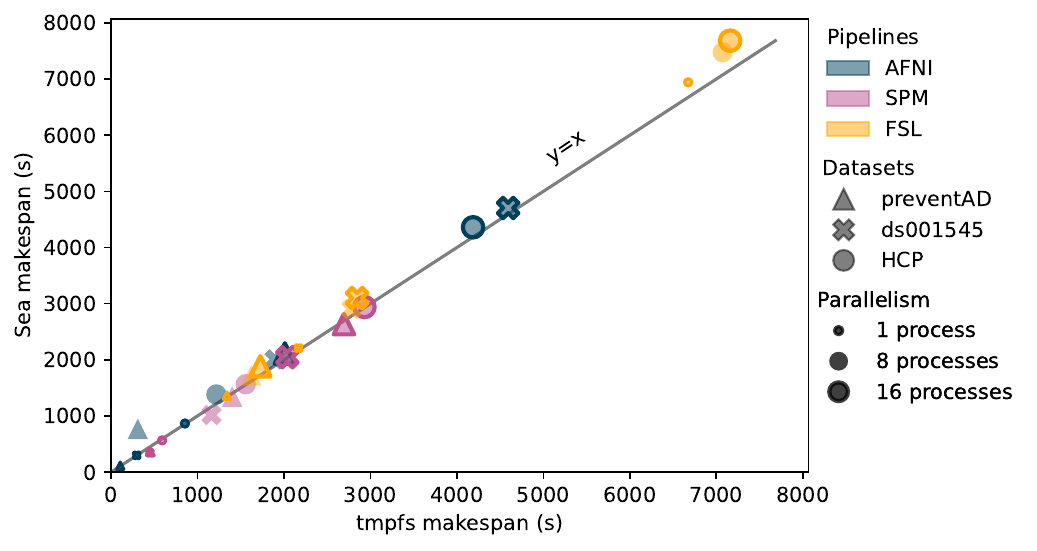}%
\caption{Makespan comparison between Sea and tmpfs on the production cluster with flushing disabled}
\label{fig:seaneuro:tmpfs}
\end{figure*}

To determine the overhead of Sea, we compared the average performance of Sea with
the performance of pipeline executing entirely within memory
(Figure~\ref{fig:seaneuro:tmpfs}). The results demonstrated that Sea and tmpfs
did not lead to significant differences in average makespan (p=0.9, two-sampled unpaired t-test). Given these results, we conclude that the
overhead of Sea is minimal and that the differences in makespan may be
due to Sea's initial reading of the data from Lustre.

\subsection{Speedups observed in production environments}

\begin{figure*}

\begin{subfigure}{0.5\textwidth}
    \centering
    \captionsetup{width=.85\linewidth}
    \includegraphics[width=\columnwidth]{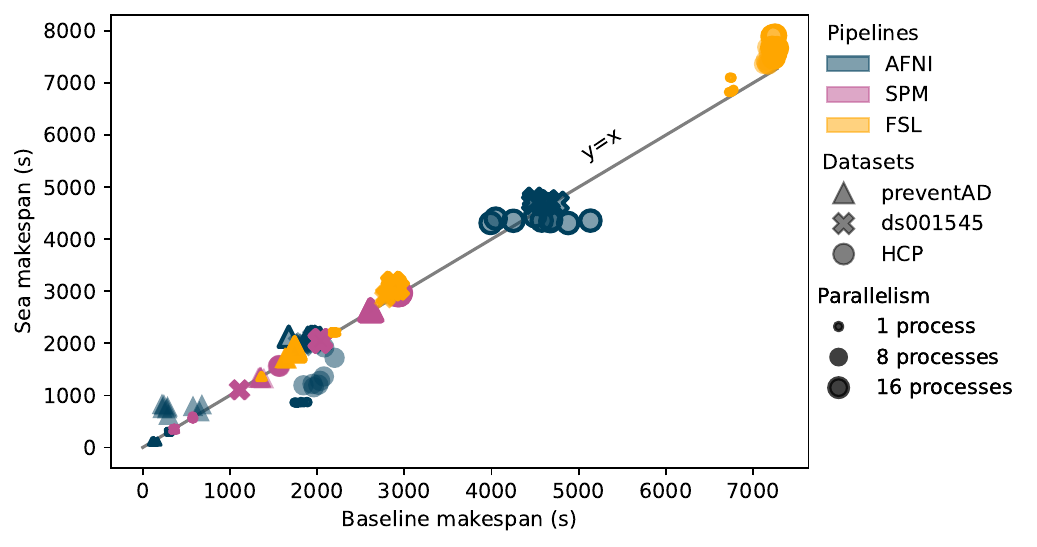}%
    \caption{Complete results}\label{fig:seaneuro:belugafull}
\end{subfigure}
\begin{subfigure}{0.5\textwidth}
    \centering
    \captionsetup{width=.85\linewidth}
    \includegraphics[width=\linewidth]{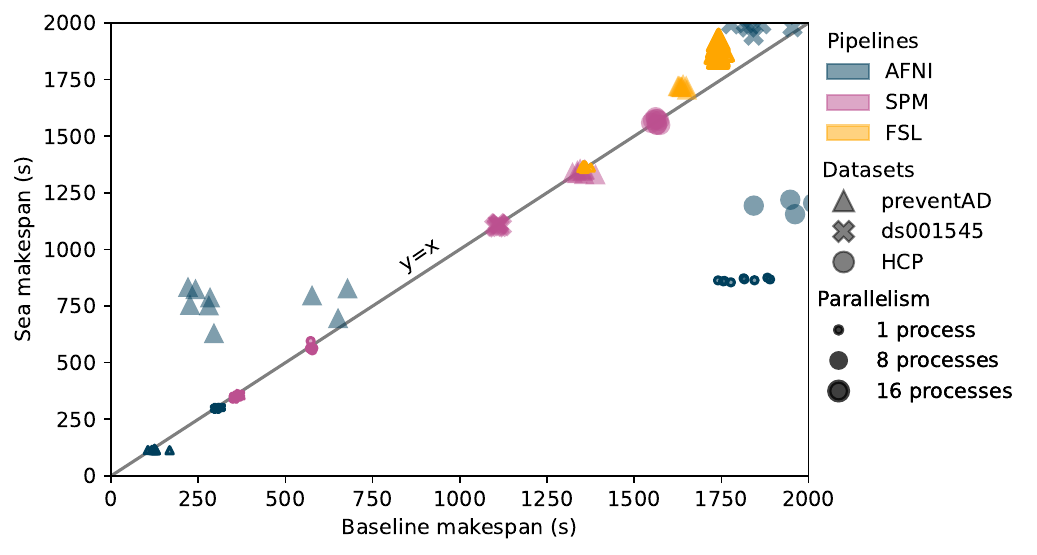}
    \caption{Zoomed between 0 and 2000 seconds}\label{fig:seaneuro:belugazoom}
\end{subfigure}
\caption{Makespan comparison between Sea and Baseline on the production cluster with flushing disabled}
\label{fig:seaneuro:beluga-noflush}
\end{figure*}

\begin{figure*}
\begin{subfigure}{0.5\textwidth}
    \centering
    \captionsetup{width=.85\linewidth}
    \includegraphics[width=\columnwidth]{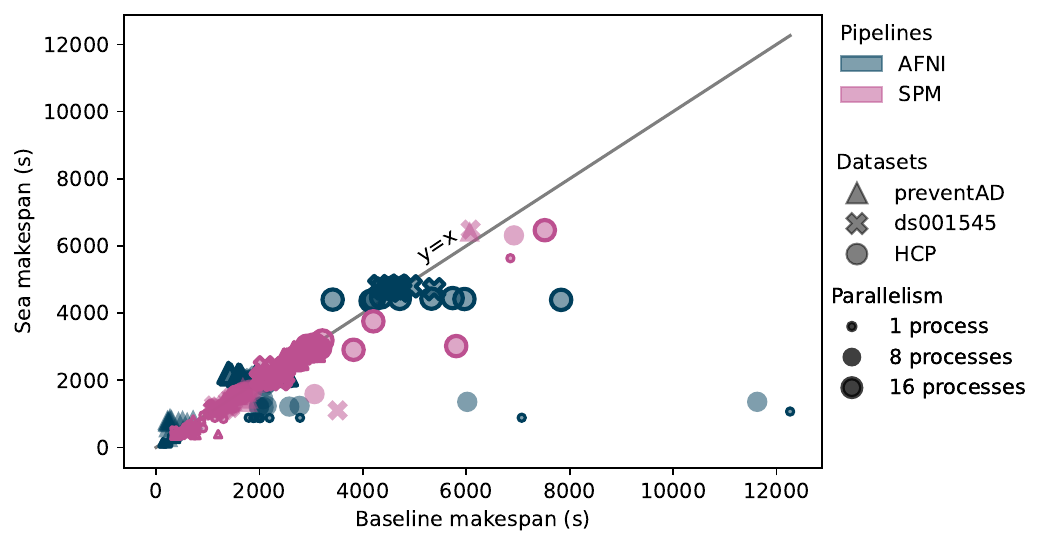}%
    \caption{Complete results}\label{fig:seaneuro:belugafullwf}
\end{subfigure}
\begin{subfigure}{0.5\textwidth}
    \centering
    \captionsetup{width=.85\linewidth}
    \includegraphics[width=\linewidth]{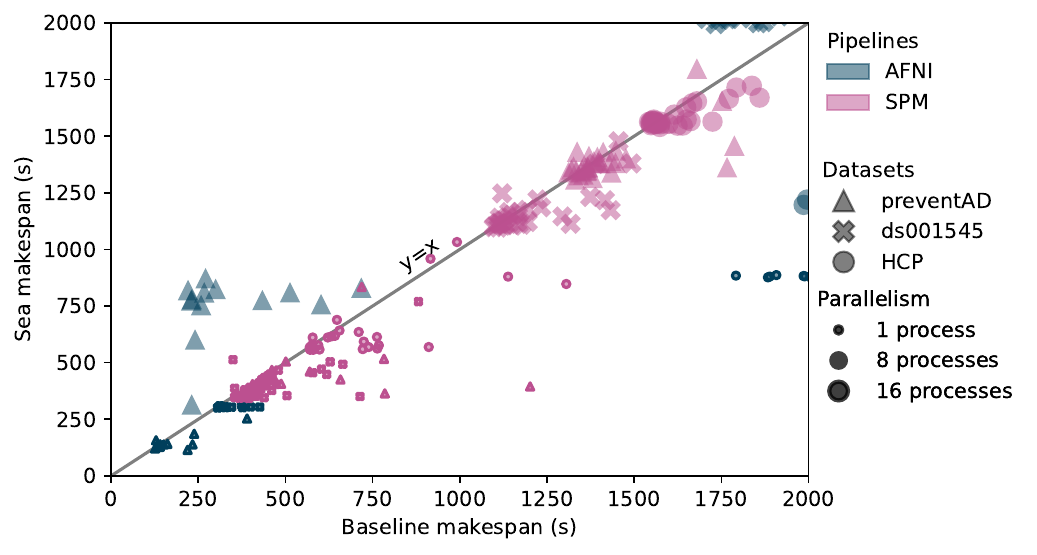}
    \caption{Zoomed between 0 and 2000 seconds}\label{fig:seaneuro:belugazoomwf}
\end{subfigure}
\caption{Makespan comparison between Sea and Baseline on the production cluster with flushing enabled for all files}
\label{fig:seaneuro:beluga-wflush}
\end{figure*}

  We also benchmarked Sea on a production cluster of the Digital Alliance of Canada
  shared with other users,
  using the same datasets and pipelines as for the controlled cluster. 
  Figures~\ref{fig:seaneuro:beluga-noflush} and \ref{fig:seaneuro:beluga-wflush}
  show that although performance in production clusters was highly variable,
  we could still obtain important speedups by using Sea. It is very likely that when we
  executed our experiments without flushing
  (Figure~\ref{fig:seaneuro:beluga-noflush}), Lustre performance was not degraded, 
  resulting in Sea and Baseline performing
  quite similarly to each other.
  
  Our results with flushing enabled confirmed this theory. When running with
  flushing enabled, which had the additional overhead of having to
  ensure that all the data was copied to Lustre, we noticed that occasionally, we
  obtained very large speedups with both the SPM and AFNI pipelines, suggesting that Lustre
  performance was degraded. The
  maximum speedup observed when running on the production cluster with flushing enabled was
  11$\times$ with the AFNI pipeline processing 1 HCP image. The maximum speedup for the SPM pipeline was
  3.2$\times$, processing 8 ds001545 images.

    \section{Discussion}
    \subsection{Significance of speedups with Sea}

    Our results show that the speedups obtained from using Sea when pre-processing functional MRI data can be 
    substantial (up to 32$\times$) depending on the degradation of the Lustre
    file system. The highest speed-ups observed on a production cluster were
    found to be 11$\times$.

    It is unrealistic for cluster users to know whether Lustre's
    performance will be deteriorated during the lifetime of their application
    execution, nor to predict how such a deterioration would impact the
    application. The fact that Sea and Baseline perform similarly in the average
    case is an indication that runtime is generally not deteriorated by Sea, as
    also demonstrated by comparing Sea's performance with tmpfs. Therefore, Sea
    is safe to use in the event where Lustre performance is sufficiently
    degraded.

    Although Sea usage did result in performance degradation at times, the magnitude
    of the degradation was less than that of possible speedups
    obtained. Our controlled experiments demonstrated that using Sea with a
    degraded Lustre file system can result in average speedups of up to
    2.5$\times$, which is likely to benefit scientific application deployments in production.

    \subsection{Limitations of using Sea}
    
    Sea's benefit is maximal when it is executed on data-intensive pipelines, that is,
     when the data is written at a rate which exceeds the rate at which the
    compute node's page cache can flush results to Lustre. In our experiments,
    the pipeline which consisted of both the shortest duration and largest
    output size was 16 AFNI pipeline processes preprocessing 16 HCP fMRI images. While the concurrent processes wrote, in total,
    approximately 100~GB of data, the controlled cluster would have had
    approximately 44~GB of dirty cache available per Lustre Object Storage Target (OST) and a maximum amount of
    100~GB of page cache. Since all writes would not have occurred at the same
    time and computation time was extended as a result of CPU contention, even
    our controlled experiments did not showcase the benefits of Sea in a highly
    data-intensive scenario. Since HCP is one of the largest fMRI datasets,
    the only way to augment data intensiveness of the applications is
    through increased parallelism which is limited by the number of available
    cores. As a result, we could only demonstrate the speedups brought by Sea when
    Lustre performance had been degraded.

    Sea cannot speedup applications that are very compute-intensive, as is the case with the FSL Feat pipeline.
    In such cases, the performance bottleneck is the
    CPU time, which Sea does not address. Moreover, depending on the
    application's I/O patterns, it is possible that there will be sufficient
    page cache space and time to flush all the results asynchronously
    to Lustre. However, even with minimal speedups, Sea can still be useful in
    these cases to limit the number of new files created on Lustre, which can be
    particularly useful in instances where users have strict file limits.

    \subsection{Predicting speedups}

    Our results expose the complexities of performance predictions on HPC clusters. We presented results on a dedicated cluster
    where we had
    control over the global cluster usage and were able to demonstrate that
    significant speedups could be obtained when all the Lustre storage devices were
    busy writing other data concurrently. However, even these experiments were
    limited in demonstrating potential speedups with Sea as there were many
    controlled variables, some of which may vary from cluster to cluster. For
    example, the production cluster had a total of 977 nodes and only 38 disks associated with
    their Lustre scratch file system. We can imagine that if 900 of the nodes
    are busy writing to the scratch file system, we would observe a speedup
    larger than what has been reported on our dedicated cluster with only 8 compute
    nodes and 44 disks associated to Lustre. However, these are not the only
    factors that can affect reported results, others can include when the
    experiment was executed, what kind of applications were being executed
    alongside the experiment, scratch file system usage during experiments, and
    what were the application read and write patterns.

    Users cannot control or predict the cluster status at the time of their
    experiments.
    Our experiments show that when parallelism is optimal (no competing threads
    and minimized compute time), we obtain speedups with Sea if Lustre
    performance is degraded. Otherwise, we should experience the same
    performance without Sea. The more data the application writes, the bigger
    the speedups, since what cannot be written to page cache can still benefit from
    being written to local storage.
    
    The frequency and the size of the user application I/O also play an
    important role in determining speedups. An application which writes a lot of
    data, but spreads out the writes evenly, may result in no speedups from
    using Sea, whereas an application which writes large amounts of data in
    bursts may benefit from significant speedups. Furthermore, applications
    which create many small files may also experience larger speedups with Sea
    as the these small files may overburden the Lustre metadata server.

    
    \subsection{Prefetching, flushing and eviction with Sea}

    Sea provides prefetching, flushing and eviction mechanisms to further
    improve the speedups that can be obtained. Prefetching allows input data to
    be moved to and manipulated from memory. All of our experiments with the SPM pipeline
    used prefetching as SPM updates the input data via the use of a memory map.
    Without prefetching, updates to the input files would have been performed
    directly on Lustre, thus exhibiting a less important speedup. It was
    favorable to proceed with prefetching, in this case, as the input files
    themselves were not large compared to available memory space. However, in
    cases with very large input files, it may be preferable to maintain the
    memory map to Lustre.

    Whereas flushing and eviction were not used in the controlled experiments,
    they were enabled for the production cluster experiments. We made the assumption, that
    users would want access to all available intermediate data, although this
    would likely not be the case with the processing of full datasets. Thus,
    eviction was mostly used to ensure that files eventually deleted by the
    pipelines would not be copied to Lustre. Slowdowns experienced with Sea may
    be a result of flushing occurring outside of the computation. We assume that
    in large scale analyses, users will not need access to all intermediate
    results, leading to better usage of tmpfs and also minimized writes to
    Lustre. Such a scenario should lead to performance results more akin to
    those obtained using Big Data frameworks.

    \subsection{Glibc interception of neuroimaging pipelines}
    
    As demonstrated by our Sea experiments, we have found that glibc interception
    is effective at capturing all file system calls executed by the
    applications. This means that Sea is compatible with three of the most
    prominently used neuroimaging preprocessing pipelines. Although we have not
    tested it extensively against all possible parameters and toolbox
    applications, we assume that use of glibc within the toolboxes themselves are
    likely to be consistent. Despite not testing against all parameters and
    toolbox applications, Sea has been extensively tested using many integration
    tests on 3 different Linux operating systems (Ubuntu, CentOS and Fedora),
    and a total of 8 different OS versions.

    The ability to use glibc interception over more common alternative methods
    such as kernel-based or FUSE-based file systems was essential in ensuring
    ease-of-use and minimal performance overheads. As both our controlled and
    production environment experiments demonstrated, Sea's overhead is minimal
    and the magnitude of the speedups to be gained by the use of Sea is
    greater than the magnitude of the slowdowns.
    
    \subsection{Should Sea always be used ?}
    
    Our results show that performance with Sea is generally hard to predict as
    users do not have control over the variables outside the application they
    are executing. We have previously discussed the conditions in which Sea may
    not be favorable to use, however, differences in the execution environment
    can still shift performance results to be favorable to Sea. As we cannot
    guarantee performance with Sea, we break down this discussion from the
    perspective of a cluster administrator and that of a user.

    From the perspective of a cluster administrator, Sea is considered useful to
    maintaining the global health of the system. By using available local
    resources, Sea is able to alleviate strain on globally shared cluster
    resources, thus improving the experience for all users. Sea is able to do
    this through the use of eviction which ensures that data that is not
    required never makes it to Lustre.

    From the perspective of a user, Sea provides speedups when
    Lustre performance is degraded and ensures limited overhead. It is most efficient
    when the compute time and non-Lustre I/O overhead is minimal. Even in
    instances where Sea does not improve performance, it helps reduce and ensure
    that the number of files that make it to Lustre are the exact ones required,
    therefore not exceeding any predetermined quota a user might have on the
    shared file system. 
    
    
    

    

    \section{Materials and Methods}

    \subsection{Datasets and Pipelines}
    
    To determine the performance gain Sea brings to neuroimaging analyses, we
    must evaluate the value of Sea on a variety of neuroimaging applications.
    For our analysis, we selected different fMRI preprocessing applications as
    fMRI processing has some of the most well-established tools for neuroimaging
    and some of the largest datasets. Of course, different modalities and tools
    may result in vastly different data access patterns and compute times.
    
    \subsubsection{Datasets}
    fMRI datasets may vary greatly in total number of images and number of
    volumes within each image. To adequately capture the diversity of datasets
    and the applicability of Sea, we have selected three datasets of varying
    time and space resolutions (Table~\ref{table:sea-neuro:data}): 1)
    OpenNeuro's ds001545 dataset~\cite{ds001545}, 2) the PREVENT-AD
    dataset~\cite{preventad}, and 3) the HCP dataset~\cite{HCP}. The ds001545
    dataset is a total of \SI{46}{\giga\byte} (1,778 files) and consists of data
    collected from 30 participants in a single session watching 3 different
    clips (i.e. three runs/participants) of The Grand Budapest Hotel.

    The PREVENT-AD dataset is an open dataset consisting of data from 330
    participants with a familial history of Alzheimer's Disease. At the time of
    the experiments, the dataset contained \SI{255}{\giga\byte} (53,061 files)
    originating from the 308 subjects available on
    DataLad~\cite{halchenko2021datalad}.
    
    The HCP project, whose aim is to characterize brain connectivity using data
    collected from 1,200 subjects, is the largest of the three datasets. At the
    time of our experiments, the dataset obtained consisted \SI{85}{\tera\byte}
    (15,716,060 files) of data collected from 1,113 subjects.
    
    By experimenting with datasets of different spatial and temporal
    resolutions, the amount of compute may be affected. Naturally, with more
    data there may inevitably be more computations performed, but also, the code
    itself may perform entirely different computations as a result of
    differences in resolution. Moreover, parallelism and program I/O behaviour
    may be impacted depending on the particular characteristics of the data.
    Capturing this potential variety in processing will provide us with a better
    understanding of when and where Sea can be useful.
    
    \begin{table*}[t]
      \small\centering
    \resizebox{\textwidth}{!}{\begin{tabular}{|c c c c c|}
      \hline
      Dataset & Total Size (MB) & Total Number of images & Number of images per experiment & Total compressed size processed (MB) \\
      \hline
      \multirow{3}{*}{PREVENT-AD} & \multirow{3}{*}{289,532} &
      \multirow{3}{*}{53,061} & 1 & 52\\
      & & & 8 & 402\\
      & & & 16 & 732\\
      \hline
      \multirow{3}{*}{ds001545} & \multirow{3}{*}{27,377} &
      \multirow{3}{*}{1,778} & 1 & 282\\
      & & & 8 & 2115\\
      & & & 16 & 4167\\
      \hline
      \multirow{3}{*}{HCP} & \multirow{3}{*}{83,140,079} &
      \multirow{3}{*}{15,716,060*} & 1 &  1301\\
      & & & 8 & 5998\\
      & & & 16 & 8328\\

      \hline
    
      \hline
    \end{tabular}} \footnotesize{*Number estimated based on data we had access
    to}\\
    \caption{Dataset characteristics. Compressed size is listed, since all
    applications, except for SPM, processed compressed
    data}\label{table:sea-neuro:data}
    \end{table*}
    
    
    
    \subsubsection{fMRI Preprocessing Pipelines}
    
    Similarly to datasets, preprocessing pipelines can also vary in duration as
    a result of methodological differences. Thus, we preprocessed each dataset
    using three standard preprocessing pipelines available within the FSL~\cite{fsl},
    AFNI~\cite{cox1996afni}, and SPM~\cite{spm} toolboxes.
    Table~\ref{table:seaneuro-pipelines} provides a reference of the differences
    in computation and data-intensiveness of the different pipelines on a single
    subject for all three of the datasets.
    
    Each tool was set to only run the functional preprocessing pipeline. For the FSL preprocessing pipeline, 
    we used FEAT configured with many of
    the default parameters. We set the following parameters: slice-timing
    correction (interleaved), intensity normalization and non-linear
    registration to standard space. FSL version 6.0.4 was used.
    
    To preprocess the datasets with SPM, we reused the template described in
    \cite{haitas2021age}, with fieldmap correction removed. As the template was
    designed for a specific dataset, we updated the paths to the fMRI and
    anatomical images, as well as the dimensions, such as the number of volumes
    and slices and the TR and TA, for each dataset. Furthermore, we set the
    slice order to interleaved in all cases. SPM version 12 was used.
    
    For AFNI preprocessing, the pipeline was configured to perform slice timing,
    alignment to Talaraich space, registration, smoothing and brain masking.
    While AFNI does have some parallelized components that can be controlled via
    environment variables, we chose to let the pipeline use as much parallelism
    as required for its execution. AFNI version 21.1.02 was used.


    
    \begin{table*}[t]
      \small\centering
      \resizebox{\textwidth}{!}{\begin{tabular}{|c c c c c c|}
      \hline
      Tool & Dataset &  Output Size (MB) & Total glibc calls & Glibc Lustre calls
      & Compute time (s)* \\
      \hline
      \multirow{3}{1em}{AFNI} & PREVENT-AD & 540 & 272,342 & 4,118 & 103.25 \\
      & ds001545 & 3,063 & 281,660 & 4,340 & 280.30 \\
      & HCP & 18,720 & 305,555 & 5,137 & 816.16 \\
      \hline
      \multirow{3}{1em}{FSL Feat} & PREVENT-AD & 254 & 191,148 & 28,099 &
      1,338.29 \\
      & ds001545 & 551 & 192,404 & 28,371 & 2,145.96 \\
      & HCP & 1,608 & 192,445 & 28,997 & 6,596.46 \\
      \hline
      \multirow{3}{1em}{SPM} & PREVENT-AD & 331.0 & 42,329 & 18,257 &  483.67 \\
      & ds001545 & 744 & 54,481 & 27,770 & 446.53 \\
      & HCP & 2,083 & 62,234 & 33,477 & 715.43 \\
    
      \hline
    \end{tabular}}
    \footnotesize{$*$ Measurements taken on the dedicated cluster}
    \caption{Pipeline execution characteristics based on the processing of a single fMRI image using a single application process}
    \label{table:seaneuro-pipelines}
  \end{table*}

    \subsection{Controls}
    Our objective was to measure the speedup brought by Sea on real neuroimaging
    pipelines. Although Sea is agnostic to the internals of the pipelines and
    their output results, incorrect parameter selection and resulting outputs
    may result in I/O patterns that would not be observed had the pipeline been
    executed properly.
   
    To validate that the outputs produced were sufficiently correct, we
    performed visual quality control of the preprocessed images to ensure that
    they were correctly skull-stripped and registered to the template brain (MNI
    template). We repeated this step for all datasets and all pipelines
    processing a single image. In addition, we compared output number of files
    and output size to ensure that the pipeline was not producing different
    outputs depending on the file system used.
    
    
    \subsection{Infrastructure}
    
    Our experiments were executed on two different clusters: 1) a controlled, dedicated cluster
    and 2) and a production cluster. We chose this experimental setup as the dedicated
    cluster allowed us to understand Sea's performance without activity of other users.
    The production cluster gave us insight into what performance typical users might
    be able to expect on a shared cluster.

    The dedicated cluster was composed of 8
    compute nodes with \SI{256}{\gibi\byte} memory and \SI{125}{\gibi\byte}
    tmpfs, running Centos 8.1.1911.
    Storage on the dedicated cluster consisted of 4 Lustre ZFS storage nodes with 44
    HDD object storage targets (OSTs) and 1 metadata server (MDS) with a single
    metadata target (MDT). The storage nodes were connected to the compute nodes
    via a 20Gbps ethernet connection. The cluster's hardware was similar to that
    of the Beluga cluster at the Digital Alliance of Canada.
    
    The Beluga cluster was used as the production cluster.
    It consisted of a total of 977 compute nodes, of which we
    used up to 16 general compute nodes at a time with \SI{186}{\gibi\byte} of
    available memory, \SI{480}{\gibi\byte} of local scratch SSD space, 2 Intel
    Gold 6148 Skylake @ 2.4 GHz CPUs and Centos 7 installed. Each compute node
    was connected to the Lustre scratch file system via an Mellanox Infiniband
    EDR (100 Gbps) network interconnect. The Lustre scratch file system consists
    of a total \SI{2.6}{\pebi\byte} of space, with 38 OSTs of
    \SI{69.8}{\tebi\byte} each and two MDTs of \SI{3.3}{\tebi\byte}.

    On the dedicated cluster, we ran the experiments with and without
    Lustre degradation, which was induced by ``busy writers''. Without busy
    writers, the fMRI preprocessing pipelines had exclusive access to the
    cluster. With busy writers, the fMRI preprocessing pipelines were executed
    alongside an Apache Spark application that continuously read and wrote approximately
    1000$\times$ \SI{617}{\mebi\byte} blocks using 64 threads, with a 5 seconds sleep between reads
    and writes. In our experiments, we either had 6 nodes each executing the
    Spark application or no busy writers.

    Our experiments generally did not use prefetching, flushing or eviction. The
    only pipeline configured to prefetch was SPM, as the input data was otherwise
    read through a memmap (i.e. only loading necessary portions of the file to memory) 
    with Lustre. Flushing would normally be necessary with
    preprocessing as the user would always require the output data. To
    investigate the impacts of flushing on these pipelines, we performed a
    separate set of experiments where AFNI and SPM would flush all data
    produced, on the production cluster.

    For all experiments, we compared the performance of Sea with that of Lustre
    alone (Baseline). To ensure that system performance was equivalent between
    Sea and our Baseline, we executed Sea and Baseline pipelines together.
    
    All experiments were executed using 1, 8 and 16 processes.
    Each process consisted of a single application call processing a single fMRI image.
    
    \section{Conclusion and Future work}

    We developed a lightweight data-management library, Sea, to seamlessly
    provide data management capabilities to existing neuroimaging workflows. We
    have tested our library on three of the most commonly used fMRI
    preprocessing libraries and found that Sea was able to intercept and
    redirect all application I/O calls without issue. 

    When benchmarking the performance of Sea compared to Baseline, it was found that Sea provided speedups
    when Lustre performance was degraded. We were able to obtain up to a
    32$\times$ in the controlled environment and 11$\times$ in a production
    cluster. It was also found that speedups were greatest with a larger
    filesize, as long as the compute time and non-Lustre I/O-related overheads
    were also minimal.

    Sea's benefits are not only limited to performance as
    users can also benefit from limiting the number of files created on Lustre
    with Sea. Archiving of the output directory on Lustre with Sea to further
    reduce number of files may be an interesting addition to Sea.

    \section{Code Availability}

    Scripts used to
    execute each pipeline can be found
    at~\url{https://github.com/valhayot/sea-slurm-scripts}. 
    Dockerfiles containing AFNI, SPM, and FSL with Sea pre-installed are available
    at~\url{https://github.com/ValHayot/Sea/tree/master/containers/applications}.

    \section{Data Availability}

    The ds001545 dataset is openly available at \url{https://openneuro.org/datasets/ds001545/versions/1.1.1} in BIDS format.
    The PREVENT-AD dataset is available at \url{https://openpreventad.loris.ca} under registered access.
    The Human Connectome Project (HCP) dataset is available at \url{https://www.humanconnectome.org} under registered access.

    \bibliography{biblio}

\begin{thebibliography}{10}
\expandafter\ifx\csname url\endcsname\relax
  \def\url#1{\texttt{#1}}\fi
\expandafter\ifx\csname urlprefix\endcsname\relax\def\urlprefix{URL }\fi
\providecommand{\bibinfo}[2]{#2}
\providecommand{\eprint}[2][]{\url{#2}}

\bibitem{gorgolewski2016brain}
\bibinfo{author}{Gorgolewski, K.~J.} \emph{et~al.}
\newblock \bibinfo{title}{The brain imaging data structure, a format for
  organizing and describing outputs of neuroimaging experiments}.
\newblock \emph{\bibinfo{journal}{Scientific data}}
  \textbf{\bibinfo{volume}{3}}, \bibinfo{pages}{1--9} (\bibinfo{year}{2016}).

\bibitem{halchenko2021datalad}
\bibinfo{author}{Halchenko, Y.~O.} \emph{et~al.}
\newblock \bibinfo{title}{Datalad: distributed system for joint management of
  code, data, and their relationship}.
\newblock \emph{\bibinfo{journal}{Journal of Open Source Software}}
  \textbf{\bibinfo{volume}{6}}, \bibinfo{pages}{3262} (\bibinfo{year}{2021}).

\bibitem{henschel2020fastsurfer}
\bibinfo{author}{Henschel, L.} \emph{et~al.}
\newblock \bibinfo{title}{Fastsurfer-a fast and accurate deep learning based
  neuroimaging pipeline}.
\newblock \emph{\bibinfo{journal}{NeuroImage}} \textbf{\bibinfo{volume}{219}},
  \bibinfo{pages}{117012} (\bibinfo{year}{2020}).

\bibitem{hoffmann2021synthmorph}
\bibinfo{author}{Hoffmann, M.} \emph{et~al.}
\newblock \bibinfo{title}{Synthmorph: learning contrast-invariant registration
  without acquired images}.
\newblock \emph{\bibinfo{journal}{IEEE transactions on medical imaging}}
  \textbf{\bibinfo{volume}{41}}, \bibinfo{pages}{543--558}
  (\bibinfo{year}{2021}).

\bibitem{HCP}
\bibinfo{author}{Van~Essen, D.~C.}, \bibinfo{author}{Smith, S.~M.},
  \bibinfo{author}{Barch, D.~M.} \emph{et~al.}
\newblock \bibinfo{title}{{The WU-Minn Human Connectome Project: an overview}}.
\newblock \emph{\bibinfo{journal}{Neuroimage}} \textbf{\bibinfo{volume}{80}},
  \bibinfo{pages}{62--79} (\bibinfo{year}{2013}).

\bibitem{ukbiobank}
\bibinfo{author}{Miller, K.~L.}, \bibinfo{author}{Alfaro-Almagro, F.},
  \bibinfo{author}{Bangerter, N.~K.} \emph{et~al.}
\newblock \bibinfo{title}{{Multimodal population brain imaging in the UK
  Biobank prospective epidemiological study}}.
\newblock \emph{\bibinfo{journal}{Nature neuroscience}}
  \textbf{\bibinfo{volume}{19}}, \bibinfo{pages}{1523} (\bibinfo{year}{2016}).

\bibitem{amunts2013bigbrain}
\bibinfo{author}{Amunts, K.} \emph{et~al.}
\newblock \bibinfo{title}{{BigBrain}: an ultrahigh-resolution {3D} human brain
  model}.
\newblock \emph{\bibinfo{journal}{Science}} \textbf{\bibinfo{volume}{340}},
  \bibinfo{pages}{1472--1475} (\bibinfo{year}{2013}).

\bibitem{zaharia2016apache}
\bibinfo{author}{Zaharia, M.} \emph{et~al.}
\newblock \bibinfo{title}{Apache {S}park: a unified engine for big data
  processing}.
\newblock \emph{\bibinfo{journal}{Comm. of the ACM}}
  \textbf{\bibinfo{volume}{59}}, \bibinfo{pages}{56--65}
  (\bibinfo{year}{2016}).

\bibitem{rocklin2015dask}
\bibinfo{author}{Rocklin, M.}
\newblock \bibinfo{title}{Dask: Parallel computation with blocked algorithms
  and task scheduling}.
\newblock In \emph{\bibinfo{booktitle}{Proc. of the 14th Python in Science
  Conference}}, \bibinfo{pages}{130--136} (\bibinfo{organization}{Citeseer},
  \bibinfo{year}{2015}).

\bibitem{rokem2021pan}
\bibinfo{author}{Rokem, A.}, \bibinfo{author}{Dichter, B.},
  \bibinfo{author}{Holdgraf, C.} \& \bibinfo{author}{Ghosh, S.}
\newblock \bibinfo{title}{Pan-neuro: interactive computing at scale with brain
  datasets} (\bibinfo{year}{2021}).

\bibitem{thunder}
\bibinfo{author}{Freeman, J.} \emph{et~al.}
\newblock \bibinfo{title}{Mapping brain activity at scale with cluster
  computing}.
\newblock \emph{\bibinfo{journal}{Nature methods}}
  \textbf{\bibinfo{volume}{11}}, \bibinfo{pages}{941} (\bibinfo{year}{2014}).

\bibitem{boubela2016big}
\bibinfo{author}{Boubela, R.~N.}, \bibinfo{author}{Kalcher, K.},
  \bibinfo{author}{Huf, W.}, \bibinfo{author}{Na{\v{s}}el, C.} \&
  \bibinfo{author}{Moser, E.}
\newblock \bibinfo{title}{Big data approaches for the analysis of large-scale
  fmri data using {A}pache {S}park and {GPU} processing: a demonstration on
  resting-state f{MRI} data from the {H}uman {C}onnectome {P}roject}.
\newblock \emph{\bibinfo{journal}{Frontiers in neuroscience}}
  \textbf{\bibinfo{volume}{9}}, \bibinfo{pages}{492} (\bibinfo{year}{2016}).

\bibitem{nipype}
\bibinfo{author}{Gorgolewski, K.} \emph{et~al.}
\newblock \bibinfo{title}{Nipype: A flexible, lightweight and extensible
  neuroimaging data processing framework in python}.
\newblock \emph{\bibinfo{journal}{Frontiers in Neuroinformatics}}
  \textbf{\bibinfo{volume}{5}}, \bibinfo{pages}{13} (\bibinfo{year}{2011}).
\newblock
  \urlprefix\url{https://www.frontiersin.org/article/10.3389/fninf.2011.00013}.

\bibitem{huang2020ooops}
\bibinfo{author}{Huang, L.} \& \bibinfo{author}{Liu, S.}
\newblock \bibinfo{title}{Ooops: an innovative tool for io workload management
  on supercomputers}.
\newblock In \emph{\bibinfo{booktitle}{2020 IEEE 26th International Conference
  on Parallel and Distributed Systems (ICPADS)}}, \bibinfo{pages}{486--493}
  (\bibinfo{organization}{IEEE}, \bibinfo{year}{2020}).

\bibitem{bb}
\bibinfo{author}{Daley, C.} \emph{et~al.}
\newblock \bibinfo{title}{Performance characterization of scientific workflows
  for the optimal use of burst buffers}.
\newblock \emph{\bibinfo{journal}{Future Generation Computer Systems}}
  \textbf{\bibinfo{volume}{110}}, \bibinfo{pages}{468--480}
  (\bibinfo{year}{2020}).
\newblock
  \urlprefix\url{https://www.sciencedirect.com/science/article/pii/S0167739X16308287}.

\bibitem{burstfs}
\bibinfo{author}{Wang, T.}, \bibinfo{author}{Mohror, K.},
  \bibinfo{author}{Moody, A.}, \bibinfo{author}{Sato, K.} \&
  \bibinfo{author}{Yu, W.}
\newblock \bibinfo{title}{An ephemeral burst-buffer file system for scientific
  applications}.
\newblock In \emph{\bibinfo{booktitle}{SC '16: Proceedings of the International
  Conference for High Performance Computing, Networking, Storage and
  Analysis}}, \bibinfo{pages}{807--818} (\bibinfo{year}{2016}).

\bibitem{gekkofs}
\bibinfo{author}{Vef, M.-A.} \emph{et~al.}
\newblock \bibinfo{title}{Gekkofs - a temporary distributed file system for hpc
  applications}.
\newblock In \emph{\bibinfo{booktitle}{2018 IEEE International Conference on
  Cluster Computing (CLUSTER)}}, \bibinfo{pages}{319--324}
  (\bibinfo{year}{2018}).

\bibitem{xtreemfs}
\bibinfo{author}{Cesario, E.} \emph{et~al.}
\newblock \bibinfo{title}{The {X}treem{FS} {A}rchitecture}.
\newblock \emph{\bibinfo{journal}{Linux {T}ag}}  (\bibinfo{year}{2007}).

\bibitem{hayot2022sea}
\bibinfo{author}{Hayot-Sasson, V.}, \bibinfo{author}{Dugr{\'e}, M.} \&
  \bibinfo{author}{Glatard, T.}
\newblock \bibinfo{title}{Sea: A lightweight data-placement library for big
  data scientific computing}.
\newblock \emph{\bibinfo{journal}{arXiv preprint arXiv:2207.01737}}
  (\bibinfo{year}{2022}).

\bibitem{ds001545}
\bibinfo{author}{Aly, M.}, \bibinfo{author}{Chen, J.},
  \bibinfo{author}{Turk-Browne, N.} \& \bibinfo{author}{Hasson, U.}
\newblock \bibinfo{title}{"learning naturalistic temporal structure in the
  posterior medial network"} (\bibinfo{year}{2019}).

\bibitem{preventad}
\bibinfo{author}{Tremblay-Mercier, J.} \emph{et~al.}
\newblock \bibinfo{title}{Open science datasets from {PREVENT-AD}, a
  longitudinal cohort of pre-symptomatic alzheimer’s disease}.
\newblock \emph{\bibinfo{journal}{NeuroImage: Clinical}}
  \textbf{\bibinfo{volume}{31}}, \bibinfo{pages}{102733}
  (\bibinfo{year}{2021}).
\newblock
  \urlprefix\url{https://www.sciencedirect.com/science/article/pii/S2213158221001777}.

\bibitem{fsl}
\bibinfo{author}{Jenkinson, M.}, \bibinfo{author}{Beckmann, C.~F.},
  \bibinfo{author}{Behrens, T.~E.}, \bibinfo{author}{Woolrich, M.~W.} \&
  \bibinfo{author}{Smith, S.~M.}
\newblock \bibinfo{title}{Fsl}.
\newblock \emph{\bibinfo{journal}{Neuroimage}} \textbf{\bibinfo{volume}{62}},
  \bibinfo{pages}{782--790} (\bibinfo{year}{2012}).

\bibitem{cox1996afni}
\bibinfo{author}{Cox, R.~W.}
\newblock \bibinfo{title}{Afni: software for analysis and visualization of
  functional magnetic resonance neuroimages}.
\newblock \emph{\bibinfo{journal}{Computers and Biomedical research}}
  \textbf{\bibinfo{volume}{29}}, \bibinfo{pages}{162--173}
  (\bibinfo{year}{1996}).

\bibitem{spm}
\bibinfo{title}{Spm}.
\newblock
  \bibinfo{howpublished}{\url{https://www.fil.ion.ucl.ac.uk/spm/software/}}.

\bibitem{haitas2021age}
\bibinfo{author}{Haitas, N.}, \bibinfo{author}{Amiri, M.},
  \bibinfo{author}{Wilson, M.}, \bibinfo{author}{Joanette, Y.} \&
  \bibinfo{author}{Steffener, J.}
\newblock \bibinfo{title}{Age-preserved semantic memory and the crunch effect
  manifested as differential semantic control networks: An fmri study}.
\newblock \emph{\bibinfo{journal}{Plos one}} \textbf{\bibinfo{volume}{16}},
  \bibinfo{pages}{e0249948} (\bibinfo{year}{2021}).

\end{thebibliography}

\end{document}